\documentclass[10pt]{article}
\usepackage{theorem}
\usepackage[mathscr]{eucal}
\usepackage{amssymb}
\newtheorem{prop}{}[section]
{\theorembodyfont{\upshape} }
\begin{document}
\hyphenation{uni-que-ness}
\newcommand{\rr}[2]{{#1}^{(#2)}}
\newcommand{\qq}[2]{{#1}^{[#2]}}
\newcommand{\boma}[1]{{\mbox{\boldmath $#1$} }}
\newcommand{\co}[2]{ {(#1)_{#2} \over #2!} }
\newcommand{\cp}[2]{ {(#1)_{#2} \over (#2)!} }
\newcommand{\Int}[1]{{#1}^{\circ}}
\newcommand{\barra}[1]{{\widetilde {#1}}}
\newcommand{\car}[1]{{\textsf{#1}}}
\def\Treves{Tr\`eves~}
\def\gen{\xi}
\def\fun{h}
\def\tfun{\tilde{\fun}}
\def\tF{\tilde{F}}
\def\tpsi{\tilde{\psi}}
\def\tf{\tilde{f}}
\def\tL{\tilde{L}}
\def\be{\beta}
\def\bei{\beta^{\dag}}
\def\ber{\beta^{\ddag}}
\def\leqs{\leqslant}
\def\geqs{\geqslant}
\def\lg{{\scriptscriptstyle{\barray{c} \vspace{-0.15cm} \leqs \\ \vspace{-0.0cm} \geqs \farray}}}
\def\supinf{{\scriptscriptstyle{\barray{c} \vspace{-0.05cm} \sup \\ \vspace{-0.0cm} \inf \farray}}}
\def\piu{{\raisebox{0.05cm}{\mbox{$\scriptscriptstyle +$}}}}
\def\meno{{\scriptscriptstyle -}}
\def\vers{\mbox{vers}}
\def\ct{\complessi^{\times}}
\def\cz{\complessi_0}
\def\ctt{\complessi^{* \times}}
\def\ctz{\ct_{\, 0}}
\def\cttz{\ctt_{~0}}
\def\Re{\mathscr R}
\def\Im{\mbox{Im}}
\def\Ro{\mathfrak R}
\def\Se{\mathscr S}
\def\Ti{\mathfrak T}
\def\implica{\Longrightarrow}
\def\sign{\mbox{sign}}
\def\di{\lambda}
\def\er{r}
\def\oa{$\overline{\mbox{a}}$}
\def\Ei{\mbox{Ei}}
\def\tu{t_1}
\def\td{t_2}
\def\tm{t_m}
\def\tM{t_M}
\def\tv{t_{\v}}
\def\Lp{L^{>}_{\v}}
\def\Lm{L^{<}_{\v}}
\def\ffi{\varphi}
\def\dfi{\delta \varphi}
\def\fip{\dot{\varphi}}
\def\dfip{\delta \dot{\varphi}}
\def\ES{{\mathscr S}}
\def\gi{{\tt g}}
\def\om{{\omega}}
\def\J{{\mathscr J}}
\def\H{{\mathscr H}}
\def\K{{\mathscr K}}
\def\Kp{{\mathscr K}'}
\def\kp{k'}
\def\scrscr{\scriptscriptstyle}
\def\scr{\scriptstyle}
\def\dd{\displaystyle}
\def\SB{\mathscr{B}}
\def\B{ B_{\mbox{\scriptsize{\textbf{C}}}} }
\def\Bc{ \overline{B}_{\mbox{\scriptsize{\textbf{C}}}} }
\def\ppartial{\overline{\partial}}
\def\d{\hat{d}}
\def\TT{{\mathcal T}}
\def\G{ {\textbf G} }
\def\Hinf{ H^{\infty}(\reali^d, \complessi) }
\def\Hn{ H^{n}(\reali^d, \complessi) }
\def\Hm{ H^{m}(\reali^d, \complessi) }
\def\Ha{ H^{\d}(\reali^d, \complessi) }
\def\Ld{L^{2}(\reali^d, \complessi)}
\def\Lpi{L^{p}(\reali^d, \complessi)}
\def\Lq{L^{q}(\reali^d, \complessi)}
\def\Lr{L^{r}(\reali^d, \complessi)}
\def\Knb{K^{best}_n}
\def\k{\mbox{{\tt k}}}
\def\x{\mbox{{\tt x}}}
\def\D{\mbox{{\tt D}}}
\def\g{ {\textbf g} }
\def\KdV{\scriptscriptstyle{KdV}}
\def\mKdV{\scriptscriptstyle{mKdV}}
\def\AKNS{\scriptscriptstyle{AKNS}}
\def\ccc{\,\makebox[0.12cm]{{$\boma{\circ} \hskip -0.190cm \circ$}}\,}
\def\QQQ{Q}
\def\XXX{X}
\def\XXXX{{\scriptstyle{{X}}}}
\def\MMMM{\scriptstyle{\MM}}
\def\MMMG{\scriptstyle{{\mathfrak{M}} }}
\def\HHH{H}
\def\WWW{W}
\def\inte{\mbox{$\int$}}
\def\iint{\inte}
\def\hhh{h}
\def\fff{f}
\def\ZZZ{\mathfrak Z}
\def\DDG{\mathfrak D}
\def\LLG{\mathfrak L}
\def\FFG{\mathfrak F}
\def\BBG{\mathfrak B}
\def\IIG{\mathfrak I}
\def\GGGG{\mathfrak G}
\def\qqq{\xi}
\def\rrr{\eta}
\def\vvv{\nu}
\def\VVV{V}
\def\MMG{\mathfrak M}
\def\VVG{\mathfrak V}
\def\WWG{\mathfrak W}
\def\RRR{R}
\def\FFF{F}
\def\MMM{M}
\def\AAA{A}
\def\III{I}
\def\BBB{B}
\def\CCC{C}
\def\GGG{G}
\def\PPP{P}
\def\DDD{\partial}
\def\gr{\mbox{graph}~}
\def\Q{$\mbox{Q}_a$~}
\def\PZ{$\mbox{P}^{0}_a$~}
\def\PZAL{$\mbox{P}^{0}_\alpha$~}
\def\PL{$\mbox{P}^{1/2}_a$~}
\def\PU{$\mbox{P}^{1}_a$~}
\def\PK{$\mbox{P}^{k}_a$~}
\def\PKU{$\mbox{P}^{k+1}_a$~}
\def\PI{$\mbox{P}^{i}_a$~}
\def\Pell{$\mbox{P}^{\ell}_a$~}
\def\PTM{$\mbox{P}^{3/2}_a$~}
\def\AZ{$\mbox{A}^{0}_r$~}
\def\AU{$\mbox{A}^{1}$~}
\def\epsilona{\epsilon^{\scriptscriptstyle{<}}}
\def\epsilonb{\epsilon^{\scriptscriptstyle{>}}}
\def\lgraffa{ \mbox{\Large $\{$ } \hskip -0.2cm}
\def\rgraffa{ \mbox{\Large $\}$ } }
\def\restriction{\upharpoonright}
\def\M{{\scriptscriptstyle{M}}}
\def\m{m}
\def\Fre{Fr\'echet~}
\def\I{{\mathcal N}}
\def\ap{{\scriptscriptstyle{ap}}}
\def\fiap{\varphi_{\ap}}
\def\EEE{ {\textbf E} }
\def\TTT{ {\textbf T} }
\def\KKK{ {\textbf K} }
\def\FFi{ {\bf \Phi} }
\def\GGam{ {\bf \Gamma} }
\def\a{\gamma}
\def\ep{\epsilon}
\def\parn{\par\noindent}
\def\teta{M}
\def\elle{L}
\def\ro{\rho}
\def\si{\sigma}
\def\ga{\gamma}
\def\de{\delta}
\def\la{\lambda}
\def\te{\vartheta}
\def\ch{\chi}
\def\et{\eta}
\def\complessi{\mathbb{C} }
\def\reali{\mathbb{R}}
\def\interi{\mathbb{Z}}
\def\naturali{\mathbb{N}}
\def\bT{{\textbf T}}
\def\T1{{\textbf T}^{1}}
\def\EE{{\mathcal E}}
\def\FF{{\mathcal F}}
\def\GG{{\mathcal G}}
\def\PP{{\mathcal P}}
\def\QQ{{\mathcal Q}}
\def\Np{{\hat{N}}}
\def\Pp{{\hat{P}}}
\def\Pip{{\hat{\Pi}}}
\def\Vp{{\hat{V}}}
\def\Ep{{\hat{E}}}
\def\Fp{{\hat{F}}}
\def\Gp{{\hat{G}}}
\def\Ip{{\hat{I}}}
\def\Tp{{\hat{T}}}
\def\Mp{{\hat{M}}}
\def\La{\Lambda}
\def\Ga{\Gamma}
\def\Si{\Sigma}
\def\Upsi{\Upsilon}
\def\Gag{{\check{\Gamma}}}
\def\Lap{{\hat{\Lambda}}}
\def\Sip{{\hat{\Sigma}}}
\def\Upsig{{\check{\Upsilon}}}
\def\Kg{{\check{K}}}
\def\ellp{{\hat{\ell}}}
\def\j{j}
\def\jp{{\hat{j}}}
\def\Stir{{\mathscr S}}
\def\M{{\mathscr M}}
\def\ess{\mathfrak s}
\def\elg{\mathfrak l}
\def\va{\mathfrak v}
\def\Ma{{\mathfrak M}}
\def\MM{{\mathcal M}}
\def\RR{{\mathcal R}}
\def\BB{{\mathcal B}}
\def\LL{{\mathcal L}}
\def\SS{{\mathcal S}}
\def\DD{{\mathcal D}}
\def\VV{{\mathcal V}}
\def\WW{{\mathcal W}}
\def\OO{{\mathcal O}}
\def\CC{{\mathcal C}}
\def\AA{{\mathcal A}}
\def\CC{{\mathcal C}}
\def\JJ{{\mathcal J}}
\def\II{{\mathcal I}}
\def\NN{{\mathcal N}}
\def\WW{{\mathcal W}}
\def\HH{{\mathcal H}}
\def\XX{{\mathcal X}}
\def\YY{{\mathcal Y}}
\def\ZZ{{\mathcal Z}}
\def\UU{{\mathcal U}}
\def\CC{{\mathcal C}}
\def\XX{{\mathcal X}}
\def\cir{{\scriptscriptstyle \circ}}
\def\circa{\thickapprox}
\def\vain{\rightarrow}
\def\ss{s}
\def\vains{\stackrel{\ss}{\rightarrow}}
\def\parn{\par \noindent}
\def\salto{\vskip 0.2truecm \noindent}
\def\spazio{\vskip 0.5truecm \noindent}
\def\vs1{\vskip 1cm \noindent}
\def\fine{\hfill $\diamond$ \vskip 0.2cm \noindent}
\newcommand{\rref}[1]{(\ref{#1})}
\def\beq{\begin{equation}}
\def\feq{\end{equation}}
\def\beqq{\begin{eqnarray}}
\def\feqq{\end{eqnarray}}
\def\barray{\begin{array}}
\def\farray{\end{array}}
\makeatletter \@addtoreset{equation}{section}
\renewcommand{\theequation}{\thesection.\arabic{equation}}
\makeatother
\begin{center}
{\huge On the Treves theorem for the AKNS equation \hskip
-0.2cm.}
\end{center}
\vspace{0.5truecm}
\begin{center}
{\large
Carlo Morosi${}^1$, Livio Pizzocchero${}^2$} \\
\vspace{0.5truecm} ${}^1$ Dipartimento di Matematica, Politecnico
di
Milano, \\ P.za L. da Vinci 32, I-20133 Milano, Italy \\
e--mail: carmor@mate.polimi.it \\
${}^2$ Dipartimento di Matematica, Universit\`a di Milano\\
Via C. Saldini 50, I-20133 Milano, Italy\\
and Istituto Nazionale di Fisica Nucleare, Sezione di Milano, Italy \\
e--mail: livio.pizzocchero@mat.unimi.it
\end{center}
\vspace{0.3truecm} {\footnotesize \textbf{Abstract.} According to
a theorem of Treves \cite{Tre2}, the conserved functionals of the AKNS
equation vanish on all pairs of formal Laurent series
$(\barra{q}, \barra{r})$ of a specified form, both of them
with a pole of the first order.
We propose a new and very simple proof for this statement,
based on the theory of B\"acklund transformations; using the
same method, we prove that the AKNS conserved functionals
vanish on other pairs of Laurent series. The spirit is
the same of our previous paper \cite{KdV} on the
Treves theorem for the KdV \cite{Tre}, with some non trivial technical
differences.
\par \vspace{0.2truecm} \noindent \textbf{Keywords:} AKNS and NLS equation, simmetries and conservation laws,
formal Laurent series.
\par \vspace{0.1truecm} \noindent \textbf{AMS 2000 Subject classifications:} 35Q53, 37K05, 37K10.}
\parn
\section{Introduction and preliminaries.}
\label{intro}
Some recent works of Treves (see \cite{Tre} \cite{Tre2} and references therein)
have introduced, in the words of Dickey \cite{Dic}, a "fresh idea" in the sector
of integrable evolutionary PDEs (KdV, nonlinear Schr\"odinger, etc.). The discovery of
Treves is that all the conserved functionals of these equations vanish
when they are evaluated on certain formal Laurent series
(intending the integrals which appear in the functionals as loop integrals in $\complessi$ around zero). \parn
To be more specific, let us consider the KdV equation $q_t = q_{x x x} - 12 q q_x$, and
the functionals $h = h(q)$ which are integrals of polynomials in $q$ and its $x$-derivatives;
for any such functional to be conserved by the KdV equation, it is necessary and
sufficient \cite{Tre} that $h(\barra{q})=0$ for all formal Laurent series with
complex coefficients of the form
$\barra{q} = 1/x^2 + \barra{q}_0 + \sum_{k=2}^{+\infty} \barra{q}_k x^k$. Now, let us pass to the coupled equations
\beq q_t = {1 \over 2} q_{x x} - q^2 r, \qquad r_t = - {1 \over 2} r_{x x} + q r^2 \label{akns} \feq
which have been considered in \cite{Tre2}; as well known, these become the nonlinear Schr\"odinger equation
$i q_{\tau} = (1/2) q_{x x} - | q |^2 q $ if $r$ is the complex conjugate of $q$
and $\tau := i t$; if $q, r$ are regarded
as independent, the pair \rref{akns} is usually called the AKNS equation
after Ablowitz, Kaup, Newell and Segur (see \cite{New} and references therein), a terminology to which we
also stick in this paper. \parn
Let us consider the
functionals $h(q,r)$ which are integrals of polynomials in $q, r$ and
their $x$ derivatives; for any such functional to be conserved by
Eq.\rref{akns}, it is necessary \cite{Tre2} that $h(\barra{q}, \barra{r})=0$ for
all pairs of Laurent series $\barra{q} = e^{\ffi} (1/x + \alpha + \beta x + \sum_{k=2}^{+\infty} \barra{\chi}_k x^k)$,
$\barra{r} = e^{-\ffi} (1/x - \alpha + \beta x + \sum_{k=2}^{+\infty} \barra{\rho}_k x^k)$
(with $\ffi$, $\alpha$, $\beta$, $\tilde{\chi}_k$, $\tilde{\rho}_k$ $\in \complessi$);
differently from the KdV case, the sufficiency of this condition has been conjectured but not proved
in \cite{Tre2}. \parn
The proofs in the original papers of Treves are based
on a long and highly technical analysis, where a central role is played by
the recurrence relations for some standard basis of conserved functionals.
A simple alternative proof of the necessary condition for the KdV, and an
analogue of this result for the Boussinesq equation, were
obtained by Dickey \cite{Dic} using the dressing method for the Lax operator.
Another proof of necessity for the KdV was proposed in a paper
of ours \cite{KdV}, where we employed the invariance of the KdV conserved functionals
under the (auto)-B\"acklund transformation. The relation between \cite{Dic} and \cite{KdV} has
been very recently discussed in \cite{Dic2}, where the extension of the B\"acklund method
to the Boussinesq and the other GD hierarchies has also been sketched. \parn
In the present paper the B\"acklund invariance of the conserved functionals will be used
in relation to the AKNS, to get a new proof of the Treves theorem
for this equation. Using this idea we will show that the AKNS conserved functionals
vanish on other nontrivial pairs of Laurent series, in particular for those
of the form $\barra{q}=
e^{\ffi} ({2/ x^2} + \alpha + \beta x + \sum_{k=2}^{+\infty} \barra{\chi}_k x^k)$,
$\barra{r} = e^{-\ffi} (1 + \beta x^3 + \sum_{k=4}^{+\infty} \barra{\rho}_k x^k)$. \parn
Both for the pairs $(\barra{q}, \barra{r})$
considered by Treves and for the ones in the above variant, our proof is very simple
and conceptually similar to the argument of \cite{KdV} for the KdV: in fact,
we show that any $(\barra{q}, \barra{r})$ as before is the B\"acklund transform of a
pair $(q,r)$ of holomorphic series (with no negative powers of $x$), on which
every functional of polynomial type is zero for trivial reasons. \parn
In comparison with our previous analysis of the KdV case, the AKNS is a bit more
difficult regarding the precise definition of the B\"acklund transformation; in fact,
even though this transformation for the AKNS is known in the literature
\cite{Cha} \cite{Tia}, its usual formulation involves rational maps
that would cause some troubles in the present framework; for this reason,
in this paper we give a different presentation which is more implicit
but involves only polynomial mappings. \parn
Let us describe the organization of the paper. In the rest of this Introduction,
we generalize for our present needs the language of differential
algebras and formal variational calculus already employed in
\cite{KdV}; the AKNS equation and the space of its conserved functionals
are described formally within this framework. In Section \ref{tret} we state precisely
the Treves theorem for the AKNS (Prop. \ref{teot}) and our variant of it mentioned
before (Prop. \ref{teov}). In Section \ref{back} we introduce the B\"acklund transformation
and state the invariance under it of the AKNS conserved functionals, in a way suitable
for our purposes; in Section \ref{newp} we use it to prove Prop.s \ref{teot} and \ref{teov}. \parn
Two Appendices have been added to review the matrix Lax formalism \textsl{\`a la} Drinfeld-Sokolov,
and its relation to B\"acklund transformations, for certain classes of integrable systems and
especially for the AKNS; the aim is, essentially, to
justify the slightly non standard presentation of the B\"acklund machinery
employed in this work.
\vskip 0.2cm \noindent
All vector spaces considered in this paper are over
$\complessi$. As anticipated, hereafter we summarise some concetps from differential algebra
and from the formal variational calculus of Gelfand-Dickey \cite{GeD}, including their applications to the AKNS
theory.
\vskip 0.2cm \noindent
\textbf{Differential algebras.}  By a differential algebra, we mean an associative algebra
(commutative or not) equipped with a derivation, i.e.,
with a linear map of the algebra into itself having the Leibnitz
property w.r.t. the product. We do not require the algebra to possess a
unity; if this exists, one easily proves that it is annihilated by the
derivation.
A morphism of differential algebras
is an algebraic morphism respecting the derivations. \parn
A differential algebra is typically written as
$(\QQ, \partial_x)$; the subscript ${~}_{x}$ attached to the derivation is also used
to denote its action on the elements $q$ of the algebra, so
$\partial_x:~ \QQ \vain \QQ$, $q \mapsto q_x$. \parn
We write $\QQ_x$ for the image of $\partial_x$; if $q, p \in \QQ$ and $p = q_x$, sometimes we say
that $q$ is a primitive of $p$. The quotient vector space
\beq \inte \QQ := \QQ/\QQ_x = \{q + \QQ_x~|~q \in \QQ \} \feq
is called the \textsl{space of integrals} of $\QQ$. The corresponding
quotient map is denoted with
\beq \inte : \QQ \vain \inte \QQ~, \qquad q \mapsto \inte q := q + \QQ_x~, \label{quot} \feq
and we call $\inte q$ the integral of $q$; of course
$\inte q_x=0$ for each $q$.
Let $\SS$ be any subset of $\QQ$; we write
\beq \inte \SS := \{s + \QQ_x~|~s \in \SS\} \subset \inte \QQ~, \label{wewrite} \feq
and note that the restriction of \rref{quot} is a map $\SS \vain \inte \SS~, s \mapsto \inte s$. \parn
Sometimes, it is necessary to specify the dependence from the differential algebra $\QQ$ of the previous operations of
integration: in this case we write $\inte^{\QQ}$ for
the map \rref{quot} or its restriction to $\SS$, and $\inte^{\QQ} \SS$ for the set \rref{wewrite}. \parn
A differential subalgebra of a differential algebra $(\QQ, \partial_x)$
is a subalgebra $\PP \subset \QQ$ closed under $\partial_x$; of course,
the differential subalgebra $\PP$ with the restricted map $\partial_x \restriction \PP \equiv \partial_x$
is itself a differential algebra. In this situation, one has to distinguish between
two kinds of integration, the first one intrinsic for the differential algebra $(\PP, \partial_x)$, and
the second one relative to $\QQ$. In the first case, we introduce as usually the
quotient space and the quotient map
\beq \inte \PP := \PP/\PP_x~; \quad \inte : \PP \mapsto \inte \PP~,~~p \mapsto \inte p := p + \PP_x~; \feq
in the second case, we consider the space and the map
\beq \inte^{\QQ} \PP := \{ p + \QQ_x~|~p \in \PP\} \subset \inte \QQ~; \qquad \inte^{\QQ} :
\PP \mapsto \inte^{\QQ} \PP~,~~p \mapsto \inte^{\QQ} p := p + \QQ_x~. \feq
One easily checks the existence of a unique map
\beq \psi : \inte \PP \vain \inte^{\QQ} \PP\quad \mbox{such that} \quad \psi (\inte p) =
\inte^{\QQ} p ~~\forall p \in \PP~. \feq
The map $\psi$ is linear and onto; furthermore, it is injective if and only if
\beq \PP \cap \QQ_x = \PP_x~. \label{ifoif} \feq
If condition \rref{ifoif} holds, we will say that $\PP$ is a \textsl{strict differential subalgebra} of $\QQ$.
(Of course, for any differential subalgebra $\PP$ it is $\PP \cap \QQ_x \supset \PP_x$;
Eq.\rref{ifoif} means that any element of $\PP$ with a primitive in $\QQ$ also has a
primitive in $\PP$). \parn
Throughout the paper, the term \textsl{ideal} is employed with the usual sense; a \textsl{differential
ideal} of a differential algebra is an ideal closed under the derivation. \parn
\vskip 0.2cm \noindent
\textbf{Gelfand-Dickey differential algebra in any number $\boma{\a}$ of generators.}
This is the commutative differential algebra
\beq \FFG  := \complessi[\gen_1,...\gen_\a,\gen_{1, x}, ..., \gen_{\a,x},...]_{0}~,
\label{fgn} \feq
made of complex polynomials in infinitely many indeterminates
$\gen_s, \gen_{s, x}$,$\gen_{s, xx}$, ... $(s=1,...,\a)$, without free term (the absence of this
is indicated by the subscript ${}_{0}$); $\FFG$ is equipped with the unique
derivation $\DDD_x \equiv {\cdot}_x$ such that
\beq (\gen_s)_x = \gen_{s, x}~~,\qquad (\gen_{s, x})_x = \gen_{s, xx}~~, \qquad ...~. \feq
We write $\FFF,\GGG$, etc. for the elements of $\FFG$. (For example:
$\FFF := \xi^2_{1, x} \xi_{2, x x}$, $\GGG := 3 \xi_{1, x} \xi_2$ $\in \FFG$;
$\FFF \GGG = 3 \xi_{1, x}^3 \, \xi_{2} \, \xi_{2, x x}$). \parn
For the elements of $\iint \FFG = \FFG/\FFG_x$, which have the form $\fff = \iint \FFF$ ($\FFF \in \FFG$),
the general denomination of "integrals" is of course available; however,
in this case the name \textsl{functionals} is more standard. \parn
The Gelfand-Dickey algebra $\FFG$ can be
represented in terms of transformations on any commutative differential algebra
$(\QQ, \partial_x)$, in the following way.
Let us consider the Cartesian product $\times^\a \QQ \equiv \QQ^\a$; then, any $\FFF \in \FFG$
induces a map
\beq F(~) : \QQ^\a \vain \QQ~, \qquad (q_1,...,q_\a) \mapsto F(q_1,...,q_\a) \label{maps} \feq
where $F(q_1,...,q_\a)$ is obtained from the expression of the polynomial $\FFF$
replacing $\gen_s$ with $q_s$, $\gen_{s, x}$ with $q_{s, x}$, etc.~. \parn
It is important to distinguish the elements of $\FFG$ from the maps on $\QQ^\a$: this is
the reason why the symbol of the map in \rref{maps} contains a bracket $(~)$. (In \cite{KdV}, for the same
reason we used bold symbols for the elements of $\FFG$, and non bold notations for the maps
on $\QQ$; in the present framework, the proliferation of bold symbols would be excessive).
We note that
\beq F(~) \in Pol(\QQ^\a, \QQ)~, \feq
where $Pol(\XX,\YY)$ are the polynomial maps \cite{KdV} between any two vector spaces $\XX,\YY$
and $\QQ^\a$ is regarded as a vector space with the product structure.
$Pol(\QQ^\a, \QQ)$ is a commutative algebra with the pointwise product, and a
differential algebra with the unique derivation $\partial_x : P(~) \mapsto P_x(~)$
such that $P_x(q) := P(q)_x$ for all $P(~)$. The correspondence
\beq \FFG \vain Pol(\QQ^\a, \QQ)~, \qquad \FFF \mapsto F(~) \feq
is a morphism of differential algebras. It also induces a linear map
\beq \iint \FFG \vain Pol(\QQ^\a, \inte \QQ)~, \qquad \fff \mapsto f(~) \feq
in the following way: if $\fff = \iint \FFF$, then
\beq f(~) : \QQ^\a \vain \inte \QQ~, \qquad (q_1,...,q_\a) \mapsto f(q_1,...,q_\a) := \inte F(q_1,...,q_\a)~.
\label{mmapps} \feq
The maps \rref{maps} and \rref{mmapps} will be called the \textsl{representations} on $\QQ$ of
$F$ and $f$, respectively.
\vskip 0.2cm \noindent
\textbf{Vector fields and Lie derivatives.} We
consider again the Gelfand-Dickey differential algebra \rref{fgn}. In this
framework, by a vector field we simply mean a family
\beq \XXX = (\XXX_1,...,\XXX_\a) \in \FFG^{\a} \feq
(this is represented as map $X(~) = (X_1(~),...,X_\a(~)) \in Pol(\QQ^\a, \QQ^\a)$
on any commutative differential algebra $\QQ$). The Lie derivative
on $\FFG$ induced by $\XXX$ is the unique derivation
\beq \LL_{\XXXX} : \FFG \vain \FFG \quad \mbox{such that} \quad
\LL_{\XXXX} \DDD_x = \DDD_x \LL_{\XXXX}~,~~\LL_{\XXXX} \gen_s = \XXX_s \quad (s=1,...,\a)~;
\label{lie1} \feq
the corresponding Lie derivative on $\iint \FFG$ is the unique map
\beq \LL_{\XXXX} : \iint \FFG \vain \iint \FFG \quad \mbox{such that} \quad \LL_{\XXXX}
\iint   = \iint \LL_{\XXXX}~; \label{lie2} \feq
of course this map is linear. The set of \textsl{conserved functionals} of a vector field $\XXX$ is
\beq \ZZZ_{\XXXX} := \{~\hhh \in \iint \FFG~|~\LL_{\XXXX} \hhh = 0 \}~; \feq
this is a vector subspace of $\iint \FFG$.
\vskip 0.2cm \noindent
\textbf{The AKNS theory and its conserved functionals} \cite{New}. This is a theory
in $\a=2$ components. For our purposes, it is necessary to formulate it in the
language of formal variational calculus; so, we introduce the Gelfand-Dickey algebra
\beq \FFG := \complessi[\qqq, \rrr, \qqq_x, \rrr_x,...]_0 \feq
with generators $\gen_1 \equiv \qqq, \gen_2 \equiv \rrr$. The AKNS vector
field is
\beq \XXX_{\AKNS} \equiv X := \big({1 \over 2} \qqq_{x x} - \qqq^2 \rrr, - {1 \over 2} \rrr_{x x} + \qqq \rrr^2 \big)
\in \FFG^2 ~.\feq
The space of conserved functionals
\beq \ZZZ_{\XXXX_{\AKNS}} \equiv \ZZZ \feq
is known to be of infinite dimension, a remarkable property placing this vector
field within the realm of integrable systems. In the Appendices
\ref{matr}, \ref{proofof} we will review the Lax formalism to obtain
the conserved functionals of this vector field and of similar systems.
This approach gives a basis $(\hhh_i)_{i=1,2,...}$ for $\ZZZ$, derived
from the "fundamental invariants" of the Lax operator: the first elements are
\beq \hhh_1 := {1 \over 2} \iint \qqq \rrr, \quad \hhh_2 := {1 \over 4} \iint \qqq_x \rrr, \quad
\hhh_3 := - {1 \over 8} \iint (\qqq^2 \rrr^2 + \qqq_x \rrr_x), \quad
\hhh_4 := {1 \over 16} \iint (-3 \rrr^2 \qqq \qqq_x + \qqq_x \rrr_{xx})~. \label{hk} \feq
\parn
\section{The Treves theorem (and some variant of it) for the AKNS.}
\label{tret}
As in the case of the KdV \cite{Tre}, this theorem concerns the representation
of an integrable system on a peculiar differential algebra. This is the
commutative differential algebra of formal Laurent series in one
indeterminate $x$ and complex coefficients, i.e.,
\beq \QQ := \{ q = \sum_{k = k_{min}}^{+\infty} q_k x^k~|~q_k \in \complessi~\forall k,~
k_{min} = k_{min}(q) \in \interi \}~; \label{spaceqq}
\feq
the product is the usual Cauchy product of series, and the derivation is
\beq \partial_x : \QQ \vain \QQ_x, \quad q \mapsto q_{x} := \sum_{k=k_{min}}^{+\infty} k q_k x^{k-1}~.
\label{deriv} \feq
Clearly, we have
\beq \QQ_x= \{ q \in \QQ~|~q_{-1} = 0 \} \feq
and the map $\inte \QQ \vain \complessi$, $\inte q \mapsto q_{-1}$
is a linear isomorphism. For this reason, from now on we make the
identifications
\beq \inte \QQ \simeq \complessi~; \qquad \inte q \simeq q_{-1} ~~ \forall q \in \QQ \feq
(in \cite{KdV}, this was presented for simplicity as the very \textsl{definition} of $\inte$).
Of course, the above description of $\mbox{$\inte q$}$ as the "residue" $q_{-1}$ suggests to
interpret it as a loop integral in $\complessi$ around zero. \parn
We come to the Treves theorem for $\ZZZ_{\AKNS} \equiv \ZZZ$ and for the differential
algebra (\ref{spaceqq}-\ref{deriv}).
In our notations, this reads:
\begin{prop}
\label{teot} \textbf{Proposition} \cite{Tre2}. Let $\hhh \in
\ZZZ \subset \iint \FFG$, and consider its representation $h(~) : \QQ^2 \vain \complessi$. Then
\beq h(\barra{q}, \barra{r})=0 \quad \forall~~\barra{q} =
e^{\ffi} \Big({1 \over x} + \alpha + \beta x + \sum_{k=2}^{+\infty} \barra{\chi}_k x^k \Big),
\quad \barra{r} =
e^{-\ffi} \Big({1 \over x} - \alpha + \beta x + \sum_{k=2}^{+\infty} \barra{\rho}_k x^k \Big),
\label{eqteo} \feq
($\ffi, \alpha, \beta, \barra{\chi}_k, \barra{\rho}_k \in \complessi)$. \fine
\end{prop}
\parn
As anticipated, our aim in this paper is to give a new proof of this result,
based on the B\"acklund transformations. This technique will allow us to prove the following
variant of the previous result:
\begin{prop}
\label{teov} \textbf{Proposition}. Let $\hhh \in \ZZZ$; then
\beq h(\barra{q}, \barra{r})=0 \quad \forall~~\barra{q} =
e^{\ffi} \Big({2 \over x^2} + \alpha + \beta x + \sum_{k=2}^{+\infty} \barra{\chi}_k x^k \Big),
\quad \barra{r} =
e^{-\ffi} \Big(1 + \beta x^3 + \sum_{k=4}^{+\infty} \barra{\rho}_k x^k \Big),
\label{eqteov} \feq
($\ffi, \alpha, \beta, \barra{\chi}_k, \barra{\rho}_k \in \complessi)$. \fine
\end{prop}
Both Prop.s \ref{teot} and \ref{teov} are proved in Section \ref{newp}. Hereafter, as
a preliminary step we discuss the AKNS B\"acklund transformation, in a formulation suitable
for our purposes.
\section{B\"acklund transformation for the AKNS theory.}
\label{back}
Essentially, this is a transformation leaving invariant the AKNS
conserved functionals. However, its description in the language of
formal variational calculus requires some technical subtleties introduced hereafter.
To this purpose, we consider besides $\FFG := \complessi[\qqq,\rrr,\qqq_x, \rrr_x,...]_0$
a "copy" of it, say
\beq \barra{\FFG} := \complessi[\barra{\qqq},\barra{\rrr},
\barra{\qqq}_x, \barra{\rrr}_x,...]_0~, \feq
with the derivation such that $(\barra{\qqq})_x = \barra{\qqq}_x$, etc.~.
Of course, there is a unique differential-algebraic isomorphism $\barra{\cdot} :
\FFG \vain \barra{\FFG}$, $\FFF \mapsto \barra{\FFF}$ sending
$\qqq, \rrr$ into $\barra{\qqq},\barra{\rrr}$. This also induces a linear isomorphism
\beq \iint \FFG \vain \iint \barra{\FFG}~, \qquad \fff = \iint \FFF \mapsto
\barra{\fff} := \iint \barra{\FFF}~. \label{tofol} \feq
We interpret $\FFG$, $\barra{\FFG}$
as describing the "initial" and "final variables" for the "transformation" to be introduced.
The latter is in fact defined implicitly in terms of an "auxiliary variable" $\vvv$;
its description mixes together the initial auxiliary and final variables, so we
introduce a third differential algebra
\beq \MMG := \complessi[\qqq,\rrr,\vvv,\barra{\qqq}, \barra{\rrr},
\qqq_x, \rrr_x, \vvv_x,\barra{\qqq}_x, \barra{\rrr}_x ...]_0~, \label{mmg} \feq
writing again $\DDD_x$ for its derivation. Up to trivial identifications, we have
\beq \FFG, ~ \barra{\FFG} \subset \MMG~; \label{incl} \feq
both $\FFG$ and $\barra{\FFG}$ are
strict differential subalgebras of $\MMG$ (in the sense of \rref{ifoif}), so
\beq \iint \FFG \simeq \iint^{\MMMG} \FFG~, ~~\iint \barra{\FFG}
\simeq \iint^{\MMMG} \barra{\FFG} ~~ \subset \iint \MMG~. \feq
\begin{prop}
\label{defbg}
\textbf{Definition.} The AKNS \textsl{B\"acklund ideal} $\IIG_{AKNS} \equiv \IIG \subset \MMG$ is the
ideal of $\MMG$ generated by the elements
\beq \III_1 := \qqq_x - \barra{\qqq}_x + \vvv( \qqq + \barra{\qqq}), \quad
\III_2 := \rrr_x - \barra{\rrr}_x + \vvv( \rrr + \barra{\rrr}), \quad \III_3 :=
\vvv_x + \qqq \rrr - \barra{\qqq} \,\,\barra{\rrr}~. \qquad \diamond \label{bbg}
\feq
\end{prop}
From standard commutative algebra,
\beq \IIG = \Big\{ \sum_{j=1}^3 \FFF_j \III_j~ \Big|~ \FFF_j \in \MMG~\forall j \Big \}~. \label{bbbg} \feq
In the language of formal variational calculus,
the B\"acklund invariance of the AKNS conserved functionals can be expressed as follows.
\begin{prop}
\label{propinv}
\textbf{Proposition.} Let $\hhh \in \ZZZ_{\AKNS} \equiv \ZZZ \subset \FFG$, and define
$\barra{\hhh}$ following \rref{tofol}. Then
\beq {~} \qquad \qquad \qquad \qquad \qquad \qquad \qquad \barra{\hhh} - \hhh \in \iint \IIG
\qquad (\iint \equiv \iint^{\MMMG})~.
\qquad \qquad \qquad \qquad \qquad \diamond \label{tesinv} \feq
\end{prop}
The above Proposition is essentially known in the literature, even though it is not usually formulated
in the language of formal variational calculus. In any case, to make the paper self contained
we propose a proof in the Appendix \ref{proofof}. \parn
In order to exemplify Eq.\rref{tesinv}, let us
consider the functionals $\hhh_i$ in Eq.\rref{hk} ($i=1,2,3,4$) and their tilded images
$\barra{\hhh}_1 = \iint \barra{\qqq} \, \barra{\rrr}$, ...; it turns out that
\beq \barra{\hhh}_1 - \hhh_1 = - {1 \over 2} \iint \III_3, \quad
\barra{\hhh}_2 - \hhh_2 = {1 \over 4} \iint (- \rrr \III_1 + \barra{\qqq} \III_2 + \vvv \III_3)~,
\label{onefinds} \feq
$$ \barra{\hhh}_3 - \hhh_3 = {1 \over 8} \iint \big( (- \barra{\rrr} \vvv + \barra{\rrr}_x) \III_1 +
(\qqq \vvv + \qqq_x) \III_2 + (\qqq \rrr + \barra{\qqq}\, \barra{\rrr} - \vvv^2) \III_3 \big)~, $$
$$ \barra{\hhh}_4 - \hhh_4 = {1 \over 16} \iint
\big(~ (\qqq \rrr ^2 - \barra{\qqq} \rrr^2 - \qqq \rrr \barra{\rrr}
 + 2 \barra{\qqq} \rrr \barra{\rrr} + \barra{\qqq} \, \barra{\rrr}^{\,2} - \rrr \vvv ^2 +
 \vvv \barra{\rrr}_x - \barra{\rrr}_{xx})  \III_1 + $$
$$ + (- \qqq \barra{\qqq} \rrr
 + \barra{\qqq}^{\,2} \rrr - 2 \barra{\qqq}^{\,2} \barra{\rrr}
  + \barra{\qqq} \vvv^2 + \vvv  \qqq_x + \qqq \vvv_x  + \barra{\qqq} \vvv_x + \qqq_{x x})  \III_2 + $$
$$  + (- \qqq \rrr \vvv  + \barra{\qqq} \rrr \vvv  + \qqq \barra{\rrr} \vvv
  - \barra{\qqq} \barra{\rrr} \vvv  + \vvv ^3 + \rrr  \qqq_x
  + \barra{\rrr}  \qqq_x  - \qqq  \rrr_x  - \barra{\qqq}  \rrr_x )  \III_3~\big), $$
We now present the consequences of the previous statements in terms of concrete differential
algebras. From now on $(\QQ, \partial_x)$ is a commutative differential algebra; so,
the generators $I_j$ in Eq.\rref{bbg} induce maps
\beq I_j(~) : \QQ^5 \vain \QQ~, \qquad (q,r,v, \barra{q}, \barra{r}) \mapsto I_j(q,r,v,\barra{q}, \barra{r})~,
\feq
$$ I_1(q,...,\barra{r}) := q_x - \barra{q}_x + v(q + \barra{q})~,
\quad I_2(q,...,\barra{r}) := r_x - \barra{r}_x + v(r + \barra{r})~,
\quad I_3(q,...,\barra{r}) := v_x + q r - \barra{q} \, \barra{r}~. $$
\begin{prop}
\label{defback}
\textbf{Definition.} Let $\QQ^2 := \QQ \times \QQ$ and consider the set $2^{\QQ^2}$ of all
subsets of $\QQ^2$. The AKNS B\"acklund transformation for $\QQ$ is the map
\beq B_{\AKNS}(~) \equiv B(~) : \QQ^2 \vain 2^{\QQ^2}~, \qquad (q,r) \mapsto B(q,r)~, \label{bq} \feq
$$ {~} \qquad B(q, r) := \{ (\barra{q}, \barra{r}) \in \QQ^2~|~\exists v \in \QQ~\mbox{s.t.}~
I_j(q,r, v, \barra{q},\barra{r})=0~ \mbox{for}~j=1,2,3~\}~. \qquad \qquad \quad \diamond $$
\end{prop}
\begin{prop}
\label{bacinv}
\textbf{Proposition.} Let $\hhh \in \ZZZ$ and consider its representation
$h(~) : \QQ^2 \vain \complessi$. For all $(q,r) \in \QQ^2$ and
$(\barra{q},\barra{r}) \in B(q,r)$, it is
\beq h(\barra{q},\barra{r}) = h(q,r)~. \feq
\end{prop}
\textbf{Proof.} From Prop. \ref{propinv} we know that $\barra{\hhh} - \hhh =
\iint \III$ for some $\III$ in the B\"acklund ideal $\IIG$. So, using the
representation $I(~) : \QQ^5 \vain \QQ$ we find
\beq h(\barra{q},\barra{r}) - h(q,r) = \inte I(q,r,v,\barra{q},\barra{r})
\qquad \forall (q,r,v,\barra{q},\barra{r}) \in \QQ^5~. \label{fora} \feq
On the other hand, by comparison with \rref{bbbg} we have
\beq I(~) = \sum_{j=1}^3 F_j(~) I_j(~) \feq
where $F_j(~) : \QQ^5 \vain \QQ$ are certain polynomial maps.
In particular, let $\barra{q}, \barra{r} \in B(q,r)$; if $v$ is as in
\rref{bq}, we have $I_j(q,r,v,\barra{q}, \barra{r}) = 0$ implying
$I(q,r,v,\barra{q}, \barra{r}) = 0$, and Eq.\rref{fora} gives
$h(\barra{q},\barra{r}) - h(q,r) = 0$. \fine
\section{Proofs of Propositions \ref{teot}, \ref{teov}.}
\label{newp}
From now on,
$(\QQ, \boma{\cdot}_x, \inte)$ is the differential algebra of formal Laurent series
described in Section \ref{tret}.
For convenience, we consider therein the differential subalgebra of "holomorphic series"
\beq \ZZ := \{ q \in \QQ~|~q = \sum_{k=0}^{+\infty} q_k x^k~\}. \feq
Trivially, we have
\begin{prop}
\label{lemmino}
\textbf{Lemma.} Consider any $\hhh \in \iint \FFG$ and its representation $h(~) : \QQ^2 \vain
\complessi$; then
\beq h(~) \restriction \ZZ^2 = 0~. \feq
\end{prop}
\textbf{Proof.} Write $\hhh = \iint \HHH$. If $q, r \in \ZZ$, it is also $H(q,r) \in \ZZ^2$,
because this is a polynomial in $q, r$ and their derivatives. Thus, the residue
$h(q,r) = \inte H(q,r)$ is zero. \fine
We consider the AKNS B\"acklund transformation $B(~) : \QQ^2 \vain 2^{\QQ^2}$ (see Def.
\ref{defback}); then, combining the previous Lemma with the B\"acklund invariance of
all the AKNS conserved functionals (Prop. \ref{bacinv}) we get
\begin{prop}
\textbf{Proposition.} Let $\hhh \in \ZZZ_{\AKNS} \equiv \ZZZ$; then
\beq {~} \qquad \qquad \qquad \qquad
h(~) \restriction B(\ZZ^2) = 0~, \qquad \quad B(\ZZ^2) := \cup_{(q,r) \in \ZZ^2} B(q,r)~. \qquad
\qquad \qquad \diamond \label{restr} \feq
\end{prop}
We now fix the attention on the set of Laurent series appearing in the Treves theorem \ref{teot}, i.e.,
\beq \TT := \Big\{ \barra{q}, \barra{r} \in \QQ^2~\Big|~\barra{q} =
e^{\ffi} \Big({1 \over x} + \alpha + \beta x + \sum_{k=2}^{+\infty} \barra{\chi}_k x^k~\Big), \label{as1}
\feq
$$ \barra{r} =
e^{-\ffi} \Big({1 \over x} - \alpha + \beta x + \sum_{k=2}^{+\infty} \barra{\rho}_k x^k~\Big),
\quad (\ffi, \alpha, \beta, \barra{\chi}_k, \barra{\rho}_k \in \complessi) \Big\}~. $$
\begin{prop}
\textbf{Lemma.} i) Let $\barra{q}, \barra{r}$ be as in Eq.\rref{as1}. Then, there are
uniquely determined series of the form
\beq q = e^{\ffi}\Big(-\alpha + \sum_{k=2}^{+\infty} \chi_k x^k~\Big), \quad r =
e^{-\ffi} \Big(\alpha + \sum_{k=2}^{+\infty} \rho_k x^k~\Big),
\quad v = - {1 \over x}  + 2 \beta x + \sum_{k=2}^{+\infty} v_k x^k~ \label{as2} \feq
such that
\beq I_j(q,r,v,\barra{q},\barra{r})=0 \qquad (j=1,2,3)~. \label{ij} \feq
ii) Statement i) implies
\beq \TT \subset B(\ZZ^2)~. \label{impl} \feq
\end{prop}
\textbf{Proof.} i) Let $q,...,\barra{r}$ be as in Eq.s (\ref{as1}-\ref{as2}), and put for brevity
$I_j \equiv I_{j}(q,r,v,\barra{q},\barra{r})$. Direct computation gives:
\beq I_1  = e^{\ffi} \sum_{k=1}^{+\infty} J_{1 k} x^k~, \quad I_2  = e^{-\ffi} \sum_{k=1}^{+\infty} J_{2 k} x^k~,
\quad I_3  = \sum_{k=1}^{+\infty} I_{3 k} x^k~; \feq
$$ J_{1 1} := \chi_2 - 3 \barra{\chi}_2 + v_2~, \quad J_{1 2} := 2 \chi_3 - 4 \barra{\chi}_3 + 2 \beta^2 + v_3 ~,
\quad J_{1 3} := 3 \chi_4 - 5 \barra{\chi}_4 + \beta (2 \chi_2 + 2 \barra{\chi}_2 + v_2) + v_4~, $$
\beq J_{1 k} := k \, \chi_{k+1} - (k+2) \barra{\chi}_{k+1} + \beta (2 \chi_{k-1} + 2 \barra{\chi}_{k-1} +
v_{k-1}) + v_{k+1} + \sum_{\ell=2}^{k-2} v_{k-\ell} (\chi_{\ell} + \barra{\chi}_{\ell}) \qquad (k \geqs 4)~; \feq
$$ J_{2 1} := \rho_2 - 3 \barra{\rho}_2 + v_2~, \quad J_{2 2} := 2 \rho_3 - 4 \barra{\rho}_3 + 2 \beta^2 + v_3 ~,
\quad J_{2 3} := 3 \rho_4 - 5 \barra{\rho}_4 + \beta (2 \rho_2 + 2 \barra{\rho}_2 + v_2) + v_4~, $$
\beq J_{2 k} := k \rho_{k+1} - (k+2) \barra{\rho}_{k+1} + \beta (2 \rho_{k-1} + 2 \barra{\rho}_{k-1} +
v_{k-1}) + v_{k+1} + \sum_{\ell=2}^{k-2} v_{k-\ell} (\rho_{\ell} + \barra{\rho}_{\ell}) \qquad (k \geqs 4)~; \feq
$$ I_{3 1} := 2 v_2 - \barra{\rho}_2 - \barra{\chi}_2~, \quad I_{3 2} := 3 v_3 -
\alpha( \rho_2 - \chi_2 + \barra{\rho}_2 - \barra{\chi}_2) - \barra{\rho}_3 - \barra{\chi}_3 - \beta^2 ~, $$
\beq I_{3 3} := 4 v_4 - \alpha( \rho_3 - \chi_3 + \barra{\rho}_3 - \barra{\chi}_3) -
\barra{\rho}_4 - \barra{\chi}_4 - \beta (\barra{\rho}_2 + \barra{\chi}_2)~, \feq
$$ I_{3 k} := (k+1) v_{k+1} - \alpha( \rho_k - \chi_k + \barra{\rho}_k - \barra{\chi}_k) -
\barra{\rho}_{k+1} - \barra{\chi}_{k+1} - \beta (\barra{\rho}_{k-1} + \barra{\chi}_{k-1}) +
\sum_{\ell=2}^{k-2} (\chi_{k - \ell} \rho_{\ell} - \barra{\chi}_{k - \ell} \barra{\rho}_{\ell})~,~~
(k \geqs 4)~.$$
We must show that the equations $J_{1 k}=0, J_{2 k}=0, I_{3 k}=0$ for all $k \geqs 1$ have uniquely
determined solutions for the coefficients $\chi_k, \rho_k, v_k$ ($k \geqs 2$). \parn
The proof is recursive: for $k=1,2,3,...$, the equation $I_{3 k} = 0$ determines
$v_{k+1}$, and inserting the result into $J_{2 k}=0$, $J_{1 k}=0$ one determines,
respectively, $\rho_{k+1}$ and $\chi_{k+1}$. \parn
ii) Let $(\barra{q}, \barra{r}) \in \TT$, and $q, r, v$ as in i). It is clear that
$(q,r) \in \ZZ^2$; Eq.\rref{ij} means $(\barra{q}, \barra{r}) \in B(q,r)$. \fine
\textbf{Proof of the Treves theorem (Prop. \ref{teot}).} Put together Eq.s \rref{restr} and \rref{impl}. \fine
In a similar way we now prove Prop.\ref{teov}, that concerns
the set
\beq \SS := \Big\{ \barra{q}, \barra{r} \in \QQ^2~\Big|~\barra{q} =
e^{\ffi} \Big({2 \over x^2} + \alpha + \beta x + \sum_{k=2}^{+\infty} \barra{\chi}_k x^k~\Big), \label{asv1}
\feq
$$ \barra{r} =
e^{-\ffi} \Big(1 + \beta x^3 + \sum_{k=4}^{+\infty} \barra{\rho}_k x^k~\Big),
\quad (\ffi, \alpha, \beta, \barra{\chi}_k, \barra{\rho}_k \in \complessi) \Big\}~; $$
everything relies on the following
\begin{prop}
\textbf{Lemma.} i) Let $\barra{q}, \barra{r}$ be as in Eq.\rref{asv1}. Then, there are
uniquely determined series of the form
\beq q = e^{\ffi} \sum_{k=2}^{+\infty} \chi_k x^k~, \quad r =
e^{-\ffi} \Big(-1 + \sum_{k=3}^{+\infty} \rho_k x^k~\Big),
\quad v = - {2 \over x}  + \alpha x + \sum_{k=2}^{+\infty} v_k x^k~ \label{asv2} \feq
such that $I_j(q,r,v,\barra{q},\barra{r})=0$ for $j=1,2,3$. \parn
ii) Statement i) implies
\beq \SS \subset B(\ZZ^2)~. \label{impl2} \feq
\end{prop}
\textbf{Proof.} i) Let $q,...,\barra{r}$ be as in Eq.s (\ref{asv1}-\ref{asv2}), and
$I_j \equiv I_{j}(q,r,v,\barra{q},\barra{r})$. Then
\beq I_1  = e^{\ffi} \sum_{k=0}^{+\infty} J_{1 k} x^k~, \quad I_2  = e^{-\ffi} \sum_{k=2}^{+\infty} J_{2 k} x^k~,
\quad I_3  = x \sum_{k=0}^{+\infty} J_{3 k} x^k~; \feq
\beq J_{1 0} := 2 v_2 - 3 \beta~, \quad J_{1 1} := 2 v_3 - 4 \barra{\chi}_2 + \alpha^2 ~,
\quad J_{1 2} := \chi_3 + 2 v_4 + \alpha v_2 - 5 \barra{\chi}_3 + \alpha \beta~, \feq
$$ J_{1 3} := 2 \chi_4 + 2 v_5 + \alpha v_3 + \beta v_2 - 6 \barra{\chi}_4 + \alpha (\chi_2 + \barra{\chi}_2)~, $$
$$ J_{1 k} := (k-1) \chi_{k+1} + 2 v_{k+2} - (k+3) \barra{\chi}_{k+1} +
\alpha v_k + \beta v_{k-1} + \alpha( \chi_{k-1} + \barra{\chi}_{k-1} )
+ \sum_{\ell=2}^{k-2} v_{k-\ell} (\chi_{\ell} + \barra{\chi}_{\ell}) \qquad (k \geqs 4)~; $$
\beq J_{2 2} := \rho_3 - 5 \beta~, \quad J_{2 3} := 2 \rho_4 - 6  \barra{\rho}_4~,
\quad J_{2 4} := 3 \rho_5 + \alpha (\rho_3 + \beta)  - 7 \barra{\rho}_5 ~, \feq
$$ J_{2 5} := 4 \rho_6 + v_2 \rho_3 + \beta v_2 - 8 \barra{\rho}_6 + \alpha (\rho_4 + \barra{\rho}_4)~, $$
$$ J_{2 k} := (k-1) \rho_{k+1} - (k+3) \barra{\rho}_{k+1} +
\beta v_{k-3} + \alpha( \rho_{k-1} + \barra{\rho}_{k-1} )
+ \sum_{\ell=3}^{k-2} v_{k-\ell} \rho_{\ell} + \sum_{\ell=4}^{k-2} v_{k - \ell} \barra{\rho}_{\ell}
\qquad (k \geqs 6)~; $$
\beq J_{3 0} := 2 v_2 - 3 \beta~, \quad J_{3 1} := 3 v_3 - \chi_2 - \barra{\chi}_2 - 2 \barra{\rho}_4~, \quad
J_{3 2} := 4 v_4 - \chi_3 - \barra{\chi}_3 - 2 \barra{\rho}_5 - \alpha \beta~, \feq
$$
J_{3 3} := 5 v_5 - \chi_4 - \barra{\chi}_4 - 2 \barra{\rho}_6 - \alpha \barra{\rho}_4 - \beta^2~, \quad
J_{3 4} := 6 v_6 - \chi_5 - \barra{\chi}_5 - 2 \barra{\rho}_7 - \alpha \barra{\rho}_5 -
\beta \barra{\rho}_4 + \chi_2 \rho_3 - \barra{\chi}_2 \beta~, $$
$$ J_{3 k} := (k+2) v_{k+2} - \chi_{k+ 1} - \barra{\chi}_{k+1} - 2 \barra{\rho}_{k+3} - \alpha \barra{\rho}_{k+1}
- \beta \barra{\chi}_{k-2} - \beta \barra{\rho}_k
+ \sum_{\ell=3}^{k-1} \chi_{k + 1 - \ell} \rho_{\ell} - \sum_{\ell=4}^{k-1}
\barra{\chi}_{k + 1 - \ell} \barra{\rho}_{\ell}~,~~
(k \geqs 5)~.$$
Again, we must show that the equations $J_{1 k}=0, J_{2 k}=0, J_{3 k}=0$ for all $k$ have uniquely
determined solutions for the coefficients $\chi_k, \rho_k, v_k$. \parn
In fact: from $J_{1 0}=0$ and $J_{1 1} = 0$ one uniquely determines $v_2$, $v_3$;
now, the equation $J_{3 0} = 0$ is automatically fulfilled, and $J_{3 1}=0$ gives $\chi_2$. At
this point, we must determine $v_{k+2}$, $\rho_{k+1}$ and $\chi_{k+1}$ for $k \geqs 2$, which is
performed recursively in the following way. From $J_{1 k}=0$ and $J_{3 k}=0$ one
computes $v_{k+2}$ and $\chi_{k+1}$; to find them, one must solve a linear
system whose matrix $\left( \barray{cc} 2 & k-1 \\ k + 2 & -1 \farray \right)$ has
determinant $-k(k+1) \neq 0$. Finally, from $J_{2 k}=0$ one gets $\rho_{k+1}$. \parn
ii) Let $(\barra{q}, \barra{r}) \in \SS$, and $q, r, v$ as in i). Then
$(q,r) \in \ZZ^2$, and statement i) means $(\barra{q}, \barra{r}) \in B(q,r)$. \fine
\textbf{Proof of Prop. \ref{teov}.} Put together Eq.s \rref{restr} and \rref{impl2}. \fine
\textbf{Remark.} In any case, the following holds:
\beq (\barra{q}, \barra{r}) \in B(\ZZ^2),~\barra{q} = \sum_{k=a}^{+\infty} \barra{q}_k x^k,
\barra{q}_a \neq 0,~~\barra{r} = \sum_{k=b}^{+\infty} \barra{r}_k x^k,
\barra{r}_b \neq 0,~~\min(a,b) < 0 \feq
$$ \implica a + b = - 2~,~~\barra{q}_a \barra{r}_b = -\min(a,b). $$
In fact, we are assuming $I_j \equiv I_j(q,r,v,\barra{q},\barra{r})=0$ for some
holomorphic $q,r \in \ZZ$ and $v \in \QQ$. To fix the ideas, let us assume $\min(a,b) = a$; then,
from $I_1=0$ we easily infer
$v = \sum_{k=-1}^{+\infty} v_k x^k$ with $v_{-1} = a$. Inserting these facts into $I_3=0$,
we obtain $a+b=-2$ and $\barra{q}_a \barra{r}_b = -a$. If $\min(a,b) = b$ we proceed similarly,
using the equations $I_2=0$ and $I_3=0$.
\parn
Of course, if $(\barra{q}, \barra{r})$ are in the subsets $\TT$ or $\SS$ we have,
respectively, $a=b=-1$ or $a=-2, b=0$. \fine
\vfill \eject \noindent
\appendix
\section{Appendix. Matrix Lax operators: the Drinfeld-Sokolov formulation.}
\label{matr}
\textbf{Some more algebra.}
Consider any differential algebra $(\UU, \partial_x \equiv \cdot_x)$,
possibly non commutative. Then, the algebra of \textsl{differential
operators with coefficients in $\UU$} is the associative (and non commutative) algebra
$Di\!f\!f(\UU)$ with
generators
\beq \partial_x, \qquad U \quad (U \in \UU)  \feq
and defining relations:
\beq \underbrace{U V}_{\mbox{\footnotesize{product in $Di\!f\!f(\UU)$}}} =
\underbrace{U V}_{\mbox{\footnotesize{product in $\UU$}}}~;
\qquad \underbrace{\partial_x U} _{\mbox{\footnotesize{product in $Di\!f\!f(\UU)$}}}=
U \partial_x + U_x~ \qquad (U, V \in \UU). \feq
Any $D \in Di\!f\!f(\UU)$ has a representation
\beq D = \sum_{k=0}^d D_k \partial_x^k \qquad (d \in \naturali, D_k \in \UU~\forall k)~,
\feq
which is unique under the condition $D_d \neq 0$ if $d \neq 0$; the unique integer $d$
determined in this way is called the \textsl{order} of the differential operator $D$.
$\UU$ can be identified with the subalgebra of $Di\!f\!f(\UU)$ made of zero order operators; if $\UU$
has unity $1$, this is also the unity of $Di\!f\!f (\UU)$ (so, the invertible zero order operators are just the
invertible elements of $\UU$).
\parn
For any vector space $\VV$, we introduce the vector space of $n \times n$ matrices
\beq Mat_{n}(\VV) = \{ \car{V} = (V_{a b})_{a,b=1,...,n}~|~ V_{a b} \in \VV~\forall \, a, b \}~; \feq
we often consider therein the supplementary subspaces $Diag_{n}(\VV)$,
$O\!f\!f_n(\VV)$ made, respectively, by the diagonal and off-diagonal matrices. \parn
If $(\VV, \DDD_x)$ is a differential algebra, $Mat_{n}(\VV)$ is a differential algebra when
equipped with the usual "row by column" product and with the "term by term" derivation $\car{V} \mapsto \car{V}_{x} :=
(V_{a b, x})$.  Up to trivial identifications, we have $\inte Mat_n(\VV) = Mat_n(\inte \VV)$ and
$\inte \car{V} = (\inte V_{a b})$.
Of course, $Diag_n(\VV)$ is a differential
subalgebra of $Mat_n(\VV)$. If $\II$ is an ideal, or a differential ($\DDD_x$-closed) ideal in $\VV$, the same occurs for
$Mat_n(\II)$ in $Mat_n(\VV)$. \parn
Consider again a vector space, now denoted for convenience with $\WW$.
We can associate to it the vector space of formal series in one indeterminate
$\lambda$
\beq \WW(\lambda) := \{ W = \sum_{i=i_{min}}^{+\infty} W_i \lambda^{-i}~|~i_{min} = i_{min}(W)
\in \interi,
W_i \in \WW \,\forall i \}~; \feq
we will often fix the attention on the subspaces
\beq \WW(\lambda)_{\leqs} := \{ \sum_{i=0}^{+\infty} W_i \lambda^{-i}~\}~, \qquad
\WW(\lambda)_{<} := \{ \sum_{i=1}^{+\infty} W_i \lambda^{-i}~\}~. \feq
If $\WW$ is a differential algebra with derivation $\partial_x \equiv \cdot_x$, then
$\WW(\lambda)$ is a differential algebra, with the usual Cauchy product of series
and the derivation $\partial_x : \WWW \mapsto \WWW_x := \sum_{i} W_{i, x} \lambda^{-i}$. Of course,
$\WW$ can be identified with the differential subalgebra of $\WW(\lambda)$ made of
series $\WWW$ with $\WWW_i = 0$ for $i \neq 0$; $\WW(\lambda)_{\leqs}$,
$\WW(\lambda)_{<}$ are also differential
subalgebras. If $\WW$ has unity $1$, this is also the unity of $\WW(\lambda)$; the subset
\beq 1 + \WW(\lambda)_{<} = \{ 1 + \sum_{i=1}^{+\infty} W_i \lambda^{-i} \} \feq
is a group with respect to the Cauchy product. \parn
To conclude these preliminaries we point out that the notation $[~,~]$, employed in
the sequel for matrices or differential operators, stands for the usual commutator.
\vskip 0.2cm \noindent
\textbf{Drinfeld-Sokolov theory.} Let us consider a commutative differential algebra
$(\FF, \DDD_x)$; to fix the ideas, on can think $\FF$ to be the algebra
\rref{fgn} with generators $\gen_s$ $(s=1,...,\a)$, but for the moment this is not necessary.
For a given $n$, we construct from it
the matrix differential algebra $Mat_n(\FF)$. In the sequel,
elements of $\iint \FF$ and
$Mat_{n}(\iint \FF)$ will be called, respectively, the \textsl{scalar} and the
$n \times n$ \textsl{matrix integrals, or functionals} of $\FF$. \parn
For technical reasons appearing in the sequel, we need the direct sum of this algebra
and the $n \times n$ matrices with complex entries, i.e.,
\beq Mat_n(\FF) \oplus Mat_n(\complessi)~, \label{dirsum} \feq
which is in an obvious way an associative algebra (the product beween
elements of $Mat_n(\FF)$ and $Mat_n(\complessi)$ is defined
again row by column); this algebra contains $Mat_n(\FF)$ as an ideal.
The derivation $\DDD_x$ of $Mat_n(\FF)$ is extended to the previous direct sum,
prescribing that it annihilates $Mat_n(\complessi)$; in this way, \rref{dirsum} is a
differential algebra and $Mat_n(\FF)$ a differential ideal of it. Of course
the unity of \rref{dirsum} is $\car{1} := diag_n(1,...,1)$.
\parn
The next step is to form the differential algebra of formal series
\beq \big( \, Mat_{n}(\FF)\oplus Mat_n(\complessi) \, \big)
(\lambda) := \{ \car{N} = \sum_{i=i_{min}}^{+\infty} \car{N}_i \lambda^{-i}~|~i_{min} \in \interi, \car{N}_i
\in Mat_{n}(\FF) \oplus Mat_n(\complessi) \,\forall i \}  \label{laurmat} \feq
(which of course contains the differential ideal $Mat_n(\FF)(\lambda)$, made of
series as above with coefficients $\car{N}_i \in Mat_n(\FF)$). The ultimate step is
the algebra
\beq \DDG_n(\FF) := Di\!f\!f\Big( \big(\, Mat_{n}(\FF)\oplus Mat_n(\complessi) \,\big)(\lambda) \Big)~, \feq
made of differential operators with coefficients in the algebra \rref{laurmat}. \parn
\begin{prop}
\textbf{Definition.} A first order, $n \times n$ \textsl{Lax operator} is
a differential operator of the form
\beq \car{L} \equiv \car{L}_{\car{A}, \car{S}} := \DDD_x - \lambda \car{A} - \car{S} \in \DDG_n(\FF)~, \feq
$$ \car{S} \in Mat_n(\FF), \qquad \car{A} = diag(a_1,...,a_n) \in Diag_n(\complessi), \qquad a_i \neq 0~
\mbox{for all $i$},~a_i \neq a_j~\mbox{for $i \neq j$}~. $$
We will write $\LLG_n(\FF)$ for the set of these operators.
\fine
\end{prop}
The first result in the Drinfeld-Sokolov theory is a diagonalization theorem for these operators.
\begin{prop}
\label{propds}
\textbf{Proposition.} Consider an operator $\car{L} = \car{L}_{\car{A}, \car{S}} \in \LLG_n(\FF)$.
Then, there is a pair of objects
\beq \car{U} = \car{1} + \sum_{i=1}^{+\infty} \car{U}_i \lambda^{-i} \in \car{1} + Mat_{n}(\FF)(\lambda)_{<}~,
\label{eqd0} \feq
\beq \car{H} = \sum_{i=0}^{+\infty} \car{H}_i \lambda^{-i} \in Diag_{n}(\FF)(\lambda)_{\leqs}~,
\feq
such that
\beq \car{L} = \car{U} \,(\DDD_x - \lambda \car{A} - \car{H}) \,\car{U}^{-1} ~; \label{eqd} \feq
moreover, the matrix functionals
\beq  \car{h}_{\car{L}, i} \equiv \car{h}_{i} := \iint \car{H}_i \qquad (i=0,1,2,...) \label{hi} \feq
are uniquely determined by $\car{L}$.
\end{prop}
\textbf{Proof.} See Section 1 of \cite{DS}. The main point is that Eq.\rref{eqd} is
equivalent to $\car{L} \car{U} = \car{U} \,(\DDD_x - \lambda \car{A} - \car{H})$, and that the expansion in powers
of $\lambda$ of both sides in this equality gives rise to recursion equations for
the sequences $(\car{H}_i)$, $(\car{U}_i)$. \fine
\begin{prop}
\textbf{Definition.} Any pair $(\car{U}, \car{H})$ as in Prop. \ref{propds}
will be called a \textsl{diagonalizing pair} for $\car{L}$. The matrix
functionals $\car{h}_{\car{L},i}$ will be called the \textsl{fundamental invariants} of $\car{L}$.
\fine
\end{prop}
\textbf{Remark.} Suppose $(\FF, \DDD_x)$ is a strict differential subalgebra of a commutative differential
algebra $(\MM,\DDD_x)$ (Eq.\rref{ifoif}; recall that
$\iint \FF \simeq \iint^{\MMMM} \FF \subset \iint \MM$).
Then, by the uniqueness statement of the previous Proposition, the diagonalizations
of $\car{L}$ as an element of $\LLG_n(\FF)$, or as an element of $\LLG_n(\MM)$, give rise
to the same fundamental invariants which belong in any case to $Diag_{n}(\iint \FF)$.
\parn
The forthcoming Proposition considers a situation of this kind; the result stated
therein clarifies the origin of the term "invariant" for the functionals $\car{h}_i$. \fine
\begin{prop}
\label{prinv}
\textbf{Proposition.} Let $(\MM, \DDD_x)$ be a commutative differential algebra
containing $\FF$ as a strict differential subalgebra; further, let $\barra{\FF}$ denote another
strict differential subalgebra of $\MM$. \parn
Consider two operators
\beq \car{L} = \car{L}_{\car{A}, \car{S}} \in \LLG_{n}(\FF)~,
\quad \barra{\car{L}} = \car{L}_{\car{A}, \barra{\car{S}}} \in \LLG_{n}(\barra{\FF})~ \feq
(with the same $\car{A}$) and their fundamental invariants $\car{h}_{\car{L}, i} \equiv
\car{h}_i \in Diag_n(\iint \FF)$,
$\car{h}_{\barra{\car{L}}, i} \equiv \barra{\car h}_i \in Diag_n(\iint \barra{\FF})$. \parn
Let $\II$ denote an ideal of $\MM$, and assume there is
\beq \car{V}  = \car{1} + \sum_{i=1}^{+\infty} \car{V}_i \lambda^{-i} \in \car{1} + Mat_{n}(\MM)(\lambda)_{<} \feq
such that
\beq \barra{\car{L}} - \car{V} \car{L} \car{V}^{-1} \in Mat_{n}(\II)(\lambda)_{\leqs}~. \label{ass} \feq
Then
\beq \barra{\car{h}}_{i} - \car{h}_i \in Diag_{n} (\iint \II) \qquad (i=0,1,2,...) \label{thes} \feq
(where, in the last equation, $\iint \equiv \iint^{\MMMM}$).
\end{prop}
In a few words: if $\car{L}, \barra{\car{L}}$ are similar up to a series with
coefficients in the ideal $Mat_n(\II)$, their fundamental invariants coincide up to
elements of $Diag_n(\iint \II)$. This result is essential for our purposes;
since our language is slightly different from
the one of \cite{DS}, it is convenient to report the \parn
\textbf{Proof of Prop. \ref{prinv}.} We choose a diagonalizing pair
$(\car{U}, \car{H})$ for $\car{L}$, so as to fulfil Eq.s (\ref{eqd0}-\ref{eqd}), and proceed in three steps. \parn
\textsl{Step 1. There are}
\beq \car{W} \in \car{1} + Mat_n(\MM)(\lambda)_{<},~~
\car{J} \in Mat_n(\II)(\lambda)_{\leqs}
\quad \mbox{\textsl{such that}} \quad
\barra{\car{L}} = \car{W} (\DDD_x - \lambda \car{A} - \car{H} - \car{J}) \car{W}^{-1}~. \label{st} \feq
In fact, assumption \rref{ass} means
$\barra{\car{L}} = \car{V} \car{L} \car{V}^{-1} + \car{F}$, where $\car{F} \in Mat_{n}(\II)(\lambda)_{\leqs}$\,;
from here and
\rref{eqd} we get
\beq \barra{\car{L}} = \car{V} \car{U} \, (\DDD_x - \lambda \car{A} - \car{H}) \, \car{U}^{-1} \car{V}^{-1} + \car{F} =
\car{V} \car{U} \, (\DDD_x - \lambda \car{A} - \car{H} + \car{U}^{-1} \car{V}^{-1} \car{F}\, \car{V} \car{U}) \, \car{U}^{-1} \car{V}^{-1}~; \feq
this gives the thesis \rref{st}, with $\car{W} := \car{V} \car{U}$ and
$\car{J} := - \car{U}^{-1} \car{V}^{-1} \car{F} \, \car{V} \car{U}$
(the relations $\car{W} \in \car{1} + Mat_n(\MM)(\lambda)_{<}$\,, $\car{J} \in Mat_n(\II)(\lambda)_{\leqs}$\, are
easily checked from the previous definitions
recalling that $\car{1} + Mat_n(\MM)(\lambda)_{<}$ is a group,
$Mat_n(\MM)(\lambda)_{\leqs}$ a subalgebra and $Mat_n(\II)$ an ideal). \parn
\textsl{Step 2. Consider the differential ideal $\GG$ of $\MM$ generated by $\II$ (i.e., the smallest
differential ideal containing $\II$). Then
$\barra{\car{L}}$, as an element of $\LLG_n(\MM)$, admits a diagonalizing
pair $(\barra{\car{U}}, \barra{\car{H}})$ with $\barra{\car{H}}$ of the form}
\beq \barra{\car{H}} = \car{H} + \car{G}~, \qquad \car{G} \in Diag_n(\GG)(\lambda)_{\leqs}~. \label{req} \feq
To prove this, we write
\beq \barra{\car{U}} = \car{W} \car{Z}, \qquad \car{W} \,\mbox{as in Step 1},\quad \car{Z} = \car{1} +
\sum_{i=1}^{+\infty} \car{Z}_i \lambda^{-i} \in
{1} + O\!f\!f_n(\GG)(\lambda)_{<}~\mbox{to be found}~, \feq
\beq \barra{\car{H}} = \car{H} + \car{G} = \sum_{i=0}^{+\infty} (\car{H}_i + \car{G}_i) \lambda^{-i}~, \qquad \car{G}_i
\in Diag_n(\GG) ~\mbox{to be found} \feq
(recall that $O\!f\!f_n$ stands for the the off-diagonal $n \times n$ matrices). Due to these
representations and to Eq.\rref{st} for $\barra{\car{L}}$,
the diagonalizing condition $\barra{\car{L}} = \barra{\car{U}} (\DDD_x - \lambda \car{A} - \barra{\car{H}})
\barra{\car{U}}^{-1}$ is fulfilled if
\beq (\DDD_x - \lambda \car{A} - \car{H} - \car{J}) \car{Z} = \car{Z}  (\DDD_x - \lambda \car{A} - \car{H} - \car{G})~. \label{last} \feq
Recalling that $\DDD_x \car{Z} = \car{Z} \DDD_x + \car{Z}_x$ and expanding the last equation in powers
of $\lambda$, we see that \rref{last} is fulfilled if
\beq - \car{G}_i + [\car{A}, \car{Z}_{i + 1}] = \car{Z}_{i, x} - \car{J}_i +
\sum_{k=1}^{i} \Big([\car{Z}_k, \car{H}_{i - k}] + \car{Z}_k \car{G}_{i -k} - \car{J}_{i-k} \car{Z}_k \Big) \qquad (i=0,1,2,...) \label{syst} \feq
(intending $\car{Z}_{0, x} := (\car{1})_x = 0$ and $\sum_{k=1}^0 := 0$).
We will show that the system \rref{syst} can be solved recursively.
To this purpose, we introduce the projections $\DD_n, \OO_n$ of $Mat_n(\MM)$ onto the supplementary subspaces
$Diag_n(\MM)$, $O\!f\!f_n(\MM)$; furthermore, recalling that $\car{A}$ is diagonal with non zero and
all different eigenvalues, we infer that
$ad_\car{A} := [\car{A}, ~\cdot~] : O\!f\!f_n(\MM) \vain O\!f\!f_n(\MM)$ is a linear
isomorphism. These remarks yield for \rref{syst} the solution
\beq \car{G}_i = \DD_n\Big( \car{J}_i - \sum_{k=1}^{i} (\car{Z}_k \car{G}_{i -k} - \car{J}_{i-k} \car{Z}_k)\Big)~,
\label{recurs} \feq
$$ \car{Z}_{i+1} = ad_{\car{A}}^{-1} \Big( \car{Z}_{i x} - \OO_n(\car{J}_i) +
\sum_{k=1}^{i}\, [\car{Z}_k, \car{H}_{i - k}] + \sum_{k=1}^i
\OO_n(\car{Z}_k  \car{G}_{i -k} - \car{J}_{i-k} \car{Z}_k) \Big)~
\qquad (i=0,1,2,...) $$
(note that $[\car{Z}_k, \car{H}_{i - k}]$ is purely off-diagonal).
From these equations, with $i=0$, one gets $\car{G}_0 = \DD_n(\car{J}_0)$, $\car{Z}_1 = - ad_{\car{A}}^{-1} \OO_n(\car{J}_0)$; subsequently,
one uses recursively Eq.s \rref{recurs} to determine at each step $\car{G}_{i}, \car{Z}_{i+1}$. The fact that
$Mat_n(\GG)$ is a differential ideal and the structure of the above equations also make evident
that $\car{G}_i, \car{Z}_{i+1}$ belong to $Mat_n(\GG)$ for all $i$; by construction these matrices are, respectively,
diagonal and off-diagonal as required. \parn
\textsl{Step 3. Conclusion of the proof.} Let us intend $\iint \equiv \iint^{\MMMM}$
(recalling the remark on $\iint \FF$ just before the statement of this Proposition, and
the analogous one for $\iint \barra{\FF}$).
Eq.\rref{req} implies
\beq \barra{\car{h}}_{i} - \car{h}_{i} = \iint \barra{\car{H}}_i - \iint \car{H}_i =
\iint \car{G}_i \in Diag_n(\iint \GG)~; \feq
the thesis \rref{thes} follows from here, and from the remark that
\beq \iint \GG = \iint \II~. \label{tobepr} \feq
To prove Eq.\rref{tobepr}, we note that any element of the differential ideal
$\GG$ has the form
\beq G = \sum_{\ell \in \Lambda} F_\ell I_\ell^{(\sigma_\ell)} \qquad (\mbox{$\Lambda$ a finite set,
$F_\ell \in \MM$, $I_{\ell} \in \II$, $\sigma_{\ell} \in \naturali$ $\forall \ell \in \Lambda$})~, \feq
where $\cdot^{(\sigma_{\ell})}$ indicates the $\sigma_{\ell}$-th power of the derivation $\cdot_x$.
Now, standard integration by parts gives
\beq {~} \qquad \iint G = \iint I~, \qquad I := \sum_{\ell \in \Lambda} (-1)^{\sigma_{\ell}}
F^{(\sigma_{\ell})}_{\ell} I_\ell \in \II~. \qquad \qquad \qquad \qquad \diamond \feq
We now review some known relations between Lax operators, their diagonalization and evolutionary problems.
From now on, we work with the Gelfand-Dickey differential algebra
\beq \FFG = \complessi[\gen_1,..., \gen_\a, \gen_{1, x},..., \gen_{\a, x},...]_0 \feq
(see Eq.\rref{fgn}). Also, we are given a vector field $\XXX = (\XXX_1,...\XXX_\a) \in \FFG^\a$;
as explained in Section \ref{intro}, there are Lie derivative operators $\LL_{\XXXX} : \FFG \vain \FFG$ and
$\LL_{\XXXX} : \iint \FFG \vain \iint \FFG$ (see Eq.s (\ref{lie1}-\ref{lie2})); these
induce "componentwisely"  maps $\LL_{\XXXX}$ of
$Mat_{n}(\FFG)$ or $Mat_n(\iint \FFG)$ into itself. Trivially, $\LL_{\XXXX}$ can be
extended to a map of $Mat_n(\FFG) \oplus Mat_n(\complessi)$ into itself,
defining it to be zero on $Mat_n(\complessi)$.
\parn
Again trivially, we extend the Lie derivative to a map $\LL_{\XXXX}$ of
$\big( \, Mat_{n}(\FFG)\oplus Mat_n(\complessi) \, \big)(\lambda)$ into itself,
setting
$\LL_{\XXXX} (\sum_{i} \car{N}_i \lambda^{-i}) := \sum_{i} (\LL_{\XXXX} \car{N}_i) \lambda^{-i}$; we finally
define $\LL_{\XXXX} : \DDG_n(\FFG) \vain \DDG_n(\FFG)$ by $\LL_{\XXXX} (\sum_{k} \car{D}_k \DDD_x^k) :=
\sum_{k} (\LL_{\XXXX} \car{D}_k) \,\DDD_x^k$. \parn
\begin{prop}
\textbf{Definition.} $\XXX$ is said to admit an $n \times n$ \textsl{Lax formulation} if there are an
operator $\car{L} = \car{L}_{\car{A}, \car{S}} \in \LLG_n(\FFG) \subset \DDG_n(\FFG)$ and a zero order operator
$\car{C} \in Mat_n(\FFG)(\lambda) \subset \DDG_n(\FFG)$ such that
\beq \LL_{\XXXX} \car{L} = [\car{L}, \car{C}\,]~. \label{lax} \feq
\end{prop}
\begin{prop}
\textbf{Proposition.} If $\XXX$ admits a Lax formulation as above, the fundamental invariants
$\car{h}_{\car{L}, i} \equiv \car{h}_{i}$ ($i=0,1,2,...$) are conserved matrix functionals for $\XXX$:
\beq \LL_{\XXXX} \car{h}_i = 0~. \feq
\end{prop}
\textbf{Proof.} \cite{DS}, Section 1. \fine
Of course, from here we get conserved scalar functionals taking all matrix elements
$h_{i, a b} \in \iint \FFG$.
\label{ds}
\section{Appendix. Lax formalism and B\"acklund transformations for the AKNS theory.}
\label{proofof}
As in Section
\ref{back}, we consider the differential algebra
\beq \FFG = \complessi[\qqq, \rrr,\qqq_x, \rrr_x,...] \feq
with two generators $\qqq, \rrr$, and the vector field
$\XXX_{\scriptscriptstyle{\!A\!K\!N\!S}} \equiv \XXX$. This is known to admit
a Lax formulation of the type \rref{lax}, with $n=2$; the matrices $\car{S},\car{A}$ of $\car{L}$, and the
matrix $\car{C}$ are given by
\beq \car{S} := \left( \barray{cc} 0 & \qqq \\ \rrr & 0 \farray \right), \quad \car{A} :=
\left( \barray{cc} 1 & 0 \\ 0 & -1 \farray \right)~, \qquad
\car{C} := {1 \over 2} \left( \barray{cc} - \qqq \rrr & \qqq_{x} \\ -\rrr_{x} & \qqq \rrr
\farray \right) + \lambda \car{S} + \lambda^2 \car{A}~. \label{sa} \feq
This operator has a diagonalizing pair $\car{U} = \car{1} + \car{U}_1/\lambda + \car{U}_2/\lambda^2 + \car{U}_3/\lambda^3+ ...$,
$\car{H} = \car{H}_0 + \car{H}_1/\lambda + \car{H}_2/\lambda^2 +  \car{H}_3/\lambda^3 + ...$, where
\beq \car{U}_1 = {1 \over 2} \left( \barray{cc} 0 & - \qqq \\  \rrr & 0 \farray \right)~,
\qquad \car{U}_2 = {1 \over 4} \left( \barray{cc} 0 & - \qqq_x \\  - \rrr_x & 0 \farray \right)~, \feq
$$ \car{U}_3 = {1 \over 8} \left( \barray{cc} 0 & - \qqq_{xx}  + \qqq^2 \rrr \\
\rrr_{x x} - \qqq \rrr^2 & 0 \farray \right)~...~, $$
$$ \car{H}_0 = 0,~~ \car{H}_1 = {1 \over 2} \left( \barray{cc} \qqq \rrr & 0 \\  0 & - \qqq \rrr \farray \right),~~
\car{H}_2 = - {1 \over 4} \left( \barray{cc} \qqq \rrr_x & 0 \\  0 & \rrr \qqq_x \farray \right),~~
\car{H}_3 = {1 \over 8} \left( \barray{cc} \qqq \rrr_{x x} - \qqq^2 \rrr^2 & 0 \\  0
& - \rrr \qqq_{x x} + \qqq^2 \rrr^2 \farray \right)~, $$
\beq \car{H}_4 := {1 \over 16} \left( \barray{cc} - \qqq \rrr_{x x x} + \qqq \qqq_x \rrr^2 + 4 \qqq^2 \rrr \rrr_x & 0 \\
0 & - \rrr \qqq_{x x x} + \rrr \rrr_x \qqq^2 + 4 \rrr^2 \qqq \qqq_x \farray \right)~, ...~. \feq
The fundamental invariants $\car{h}_i := \iint \car{H}_i$ are conserved functionals for
$\XXX$; for all $i$, the diagonal elements of $\car{H}_i$ are opposite up to total derivatives, so that
\beq \car{h}_{i} \equiv \car{h}_{\car{L}, i}  = \left( \barray{cc} \hhh_i & 0 \\ 0 & - \hhh_i \farray \right)~.  \label{so} \feq
For $i=1,2,3,4$, the conserved scalar functionals $\hhh_i \in \iint \FFG$ are
the ones appearing in Eq.\rref{hk}. \parn
Up to a rescaling of each
element by a suitable constant, the sequence $(\hhh_i)_{i=1,2,...}$ can be
identified with the basis of
$\ZZZ_{\AKNS} \equiv \ZZZ$ considered in \cite{Tre2}. (The basis in
\cite{Tre2} is not defined via the Lax formalism but, rather, by a precise formulation
of the known biHamiltonian recursion scheme for the AKNS \cite{Mag}; the fact that the functionals
$(\hhh_i)$ fulfill this recursion scheme reflect a general feature of the Drinfeld-Sokolov
approach: see again \cite{DS}, Section 1 where a general construction is given for the biHamiltonian structure
of the evolution equations arising from matrix first order Lax operators).
\parn
We come to the B\"acklund transformation. Let us recall the
formalism of Section \ref{back} about it: this envolves the "initial variables"
$\qqq, \rrr$ generating $\FFG$, the "final variables" $\barra{\qqq}, \barra{\rrr}$
generating $\barra{\FFG}$, and the "auxiliary variable" $\vvv$ generating
with all the previous ones the differential algebra $\MMG$ (see Eq.\rref{mmg}).
Of course: the "tilde" map $\barra{\cdot} : \FFG \vain \barra{\FFG}$ induces
componentwisely a tilde map $\barra{\cdot} : Mat_n(\FFG) \vain Mat_n(\barra{\FFG})$;
the inclusions $\FFG, \barra{\FFG} \subset \MMG$ induce
inclusions of the corresponding spaces of matrices, formal series in $\lambda$ and
differential operators. In particular, we fix the attention on the Lax operators
\beq \car{L}_{\car{A},\car{S}} \equiv \car{L} \in \DDG_{2}(\FFG) \subset \DDG_2(\MMG)~, \qquad
\car{L}_{\car{A},\barra{\car{S}}} \equiv \barra{\car{L}} \in
\DDG_{2}(\barra{\FFG}) \subset \DDG_2(\MMG) \feq
($\car{A},\car{S}$ as in Eq.\rref{sa}; as stipulated before, $\barra{\car{S}}$ means
$\left( \barray{cc} 0 & \barra{\qqq} \\ \barra{\rrr} & 0 \farray \right)$)~.
\begin{prop}
\label{lem1}
\textbf{Lemma.} Let
\beq \car{V} := \car{1} + {1 \over 2 \lambda} \left( \barray{cc} \vvv  & \qqq - \barra{\qqq} \\
\barra{\rrr} - \rrr & - \vvv \farray \right)~. \feq
Then
\beq \barra{\car{L}}  - \car{V} \car{L} \car{V}^{-1} \in Mat_{2}(\IIG)(\lambda)_{<}~, \feq
where $\IIG$ is the B\"acklund ideal of Def. \ref{defbg}.
\end{prop}
\textbf{Proof.}
One finds by direct computation that
\beq \barra{\car{L}} \car{V} - \car{V} \car{L} = \car{V}_x - \barra{\car{S}} \car{V} + \car{V} \car{S} - \lambda [\car{A}, \car{V}] =
\car{I}~, \qquad  \car{I} := {1 \over 2 \lambda} \left( \barray{cc} \III_3  & \III_1 \\
- \III_2 & - \III_3 \farray \right)~, \feq
where $\III_j$ ($j=1,2,3$) are the generators \rref{bbg} of the B\"acklund ideal. This implies
$\barra{\car{L}}  - \car{V} \car{L} \car{V}^{-1} = \car{F}$,
where $\car{F} := \car{I} \car{V}^{-1} \in Mat_{2}(\IIG)(\lambda)_{<}$\,. \fine
The previous Lemma allows to give the \parn
\textbf{Proof of Prop. \ref{propinv}.}
The vector space $\ZZZ$ is generated by the sequence $(\hhh_i)$, so it suffices to prove that
\beq \barra{\hhh}_i - \hhh_i \in \iint \IIG \qquad (i=1,2,3,...)~. \feq
From Eq.\rref{so} and its tilded analogue, we know that
\beq \left( \barray{cc} \hhh_i & 0 \\ 0 & - \hhh_i \farray \right)~, \qquad
\left( \barray{cc} \barra{\hhh}_i & 0 \\ 0 & - \barra{\hhh}_i \farray \right)
\feq
are the invariant matrix functionals of $\car{L}$ and $\barra{\car{L}}$, respectively;
by Prop. \ref{prinv} and  Lemma \ref{lem1}, these differ by elements of $Diag_2(\iint \IIG)$,
yielding the thesis. \fine
As an example, in Eq.\rref{onefinds} we have given explicit representations
of $\barra{\hhh}_i - \hhh_i$ as integrals of elements of $\IIG$, for
$i=1,2,3,4$; these representations have been computed specializing to this case the
general argument employed to prove Prop. \ref{prinv}. In particular, this has required
the application of the recursion relations \rref{recurs} up to $i=4$.
\vskip 0.6cm \noindent \textbf{Acknowledgments.} This work was
partly supported by INDAM, Gruppo Nazionale per la Fisica
Matematica. 
\end{document}